\documentclass[letterpaper]{jpconf}
\usepackage{graphicx}
\usepackage{relsize}
\usepackage{xspace}

\def\babar{\mbox{\slshape B\kern-0.1em{\smaller A}\kern-0.1em B\kern-0.1em{\smaller A\kern-0.2em R}}}
\def\Bbar    {\kern 0.18em\overline{\kern -0.18em B}{}\xspace}
\def\BB      {\ensuremath{B\Bbar}\xspace} 
\newcommand{\timesix}{\ensuremath{\times10^{6}}}

\begin{document}

\title{CKM angles and sides from \babar}

\author{Alessandro Gaz}

\address{University of Colorado 
\newline 
\textit{on behalf of the \babar\ Collaboration}}

\begin{abstract}
The $CKM$ paradigm has been proven to be successful 
in explaining the flavour structure of the standard model 
and the non-trivial imaginary phase of the $CKM$ matrix is
the only know source of $CP$-violation.
$B$-meson decays allow us to precisely determine the fundamental
parameters of the $CKM$ matrix and put stringent constraints 
on the models of New Physics.
I present some of the most recent measurements related
to the $CKM$ Unitarity Triangle performed by the \babar\
experiment, located at the SLAC National Accelerator
Laboratory. Most results are based on the final \babar\ 
dataset, consisting of 467 \timesix\ \BB\ pairs.
\end{abstract}

\section{Introduction}

In the standard model, the $CKM$ matrix \cite{CKM} describes
the couplings, through charged weak currents, of up-type quarks
with down-type quarks. The $3 \times 3$ unitary matrix is determined
by four parameters: three of those can be interpreted as mixing angles
between the three pairs of generations, while the fourth parameter
is a non trivial complex phase which is the only known source of
$CP$ violation in the standard model.

The following among the unitarity constraints of the $CKM$ matrix:
\begin{equation}
V_{ud}V^*_{ub} + V_{cd}V^*_{cb} + V_{td}V^*_{tb} = 0
\end{equation}
can be used to construct a triangle in the complex plane, the so-called
Unitarity Triangle. One of the sides of the triangle has unitary length
by construction, whereas the others have lengths:
\begin{equation}
\begin{array}{c c}
\displaystyle R_u \equiv \left| \frac{V_{ud}V^*_{ub}}{V_{cd}V^*_{cb}} \right| \, , &
\displaystyle R_t \equiv \left| \frac{V_{td}V^*_{tb}}{V_{cd}V^*_{cb}} \right| \, . \\
\end{array}
\end{equation}
The angles are defined as:
\begin{equation}
\begin{array}{c c c}
\displaystyle \alpha \equiv \arg \left[ -\frac{V_{td}V^*_{tb}}{V_{ud}V^*_{ub}} \right] \, , &
\displaystyle \beta \equiv \arg \left[ -\frac{V_{cd}V^*_{cb}}{V_{td}V^*_{tb}} \right] \, , &
\displaystyle \gamma \equiv \arg \left[ -\frac{V_{ud}V^*_{ub}}{V_{cd}V^*_{cb}} \right] \, . 
\end{array}
\end{equation}
In the following, a few of the most recent measurements performed by the
\babar\ Collaboration, relevant for the determination of the elements of the
$CKM$ matrix, will be presented. Most of those are based on the full $\Upsilon(4s)$ 
dataset available to the experiment, consisting of 467 \timesix\ \BB\ 
pairs.

\section{$CKM$ sides}

\begin{figure}[htbp]
\begin{center}
\begin{tabular}{c c}
\includegraphics[height=5cm]{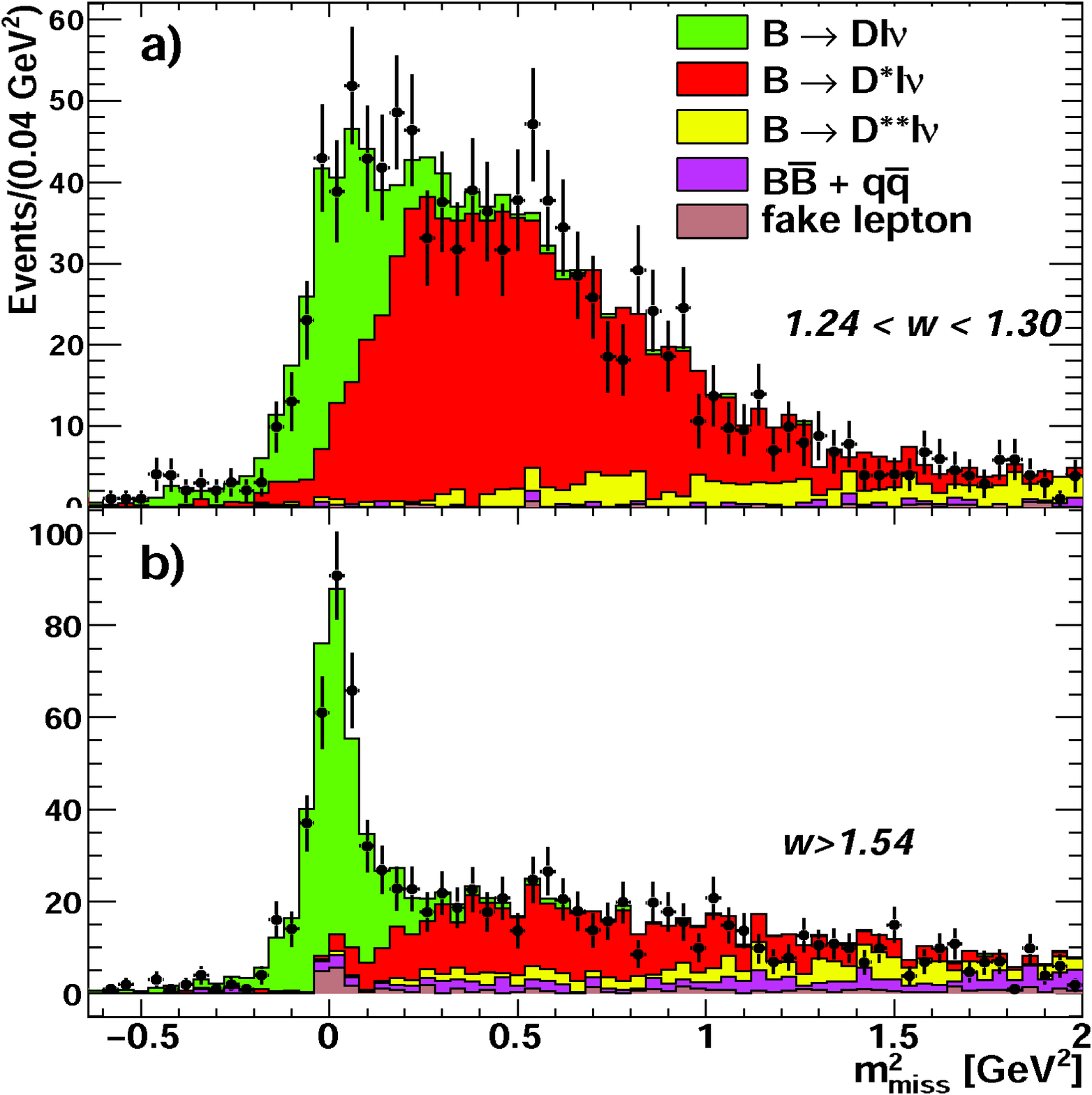} &
\includegraphics[height=5cm]{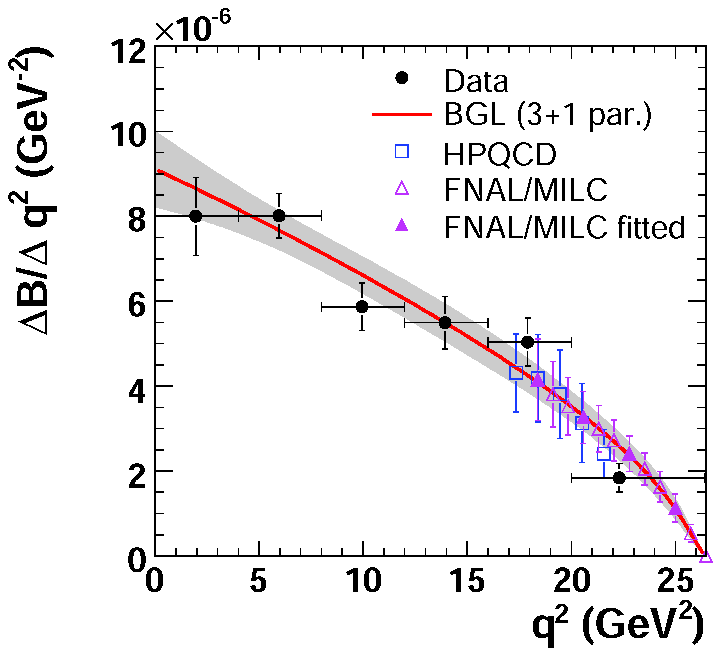} 
\end{tabular}
\caption{\label{fig:sides}Left plot: missing mass squared for the 
$B \rightarrow D \ell \nu$ analysis in two different bins of $w$.
Right plot: simultaneous fit of data and theoretical predictions
for the extraction of $|V_{ub}|$ from the branching fractions of
$B \rightarrow \pi(\rho) \ell \nu$. The results on $|V_{ub}|$ are
preliminary.}
\end{center}
\end{figure}

The element $|V_{cb}|$ can be extracted from the measurement of the
branching fraction of $B \rightarrow D \ell \nu$ decays \cite{vcb}. These 
decays are searched for on a sample where one of the two $B$'s is fully 
reconstructed in one of many hadronic final states. The measurement
of the branching fractions is performed in bins of $w$, where $w$ is
the product of the four-velocities of the $B$ and $D$ mesons. The signal
yield is extracted from a maximum likelihood fit to the missing mass
squared of the unreconstructed $B$ candidate, which peaks at zero for
signal decays (see Fig.~\ref{fig:sides}). The measured branching fractions 
are:
\begin{equation}
\begin{array}{c c}
{\cal B}(B^- \rightarrow D^0 \ell^- \nu) = (2.31 \pm 0.08 \pm 0.09)\% \, , &
{\cal B}(B^0 \rightarrow D^+ \ell^- \nu) = (2.23 \pm 0.11 \pm 0.11)\% \, ,
\end{array}
\end{equation}
where the first error is statistical and the second systematic.
From this, using the calculations from Unquenched Lattice QCD \cite{lqcd},
the value of $|V_{cb}|$ is extracted:
\begin{equation}
|V_{cb}| = (39.8 \pm 1.8 \pm 1.3 \pm 0.9) \times 10^{-3} \, , 
\end{equation}
where the first error is statistical, the second is the experimental 
systematic, and the third is the error from the theory.

In a similar way, $|V_{ub}|$ is extracted from the measurement of the
branching fractions of $B \rightarrow \pi(\rho) \ell \nu$ \cite{vub}. 
These decays are searched for inferring from the missing energy and
momentum of the event the energy and momentum of the unreconstructed
neutrino. We measure the branching fraction of the four (charged and
neutral $\pi$ and $\rho$) modes with a simultaneous maximum likelihood
fit, imposing the conservation of the isospin for the $\pi$ and $\rho$
channels. The results are:
\begin{eqnarray}
{\cal B}(B^0 \rightarrow \pi^- \ell^+ \nu) & = & (1.41 \pm 0.05 \pm 0.07) \times 10^{-4} \, , \\
{\cal B}(B^0 \rightarrow \rho^- \ell^+ \nu) & = & (1.75 \pm 0.15 \pm 0.27) \times 10^{-4} \, ,
\end{eqnarray}
where the first quoted error is statistical and the second systematic. 
Several methods can be employed to extract $|V_{ub}|$; theoretical
predictions from Lattice QCD or Light Cone Sum Rules on the form 
factor of the decays can be used, integrating part of the $q^2$
spectrum ($q$ is the four-momentum of the virtual $W$ boson exchanged
in the decay), or, following an innovative approach, experimental
data and theoretical predictions can be fitted in a simultaneous
fit (see Fig.~\ref{fig:sides}). Using the results from the FNAL and 
MILC Collaboration \cite{milc} we obtain:
\begin{equation}
|V_{ub}| = (2.95 \pm 0.31) \times 10^{-3} \, .
\end{equation}

\section{$CKM$ angles}

Information about the angles of the Unitarity Triangle can be obtained
by looking for several different $CP$ violating phenomena in $B$ decays.

Concerning $\gamma$, the \babar\ Collaboration recently presented
some results based on the measurement of the branching fractions and
charge asymmetries of $B^- \rightarrow D^{(*)0} K^{(*)-}$ decays. In
\cite{glw}, the GLW method \cite{glw_th} is used to get some non-trivial
constraints on the angle $\gamma$ from the $B^- \rightarrow D^0 K^{*-}$
decays. Analogous constraints are obtained from the ADS method \cite{ads_th}
applied to $B^- \rightarrow D^{(*)0} K^-$ decays, where the first evidence
for an ADS signal is seen \cite{ads}. Though useful, these results exclude
at the 95\% C.L. only a small range and are not competitive with the 
extraction of $\gamma$ exploiting a Dalitz Plot analysis of the $D^0$ to 
self-conjugate states.

\begin{figure}[htbp]
\begin{center}
\begin{tabular}{c c}
\includegraphics[height=5cm]{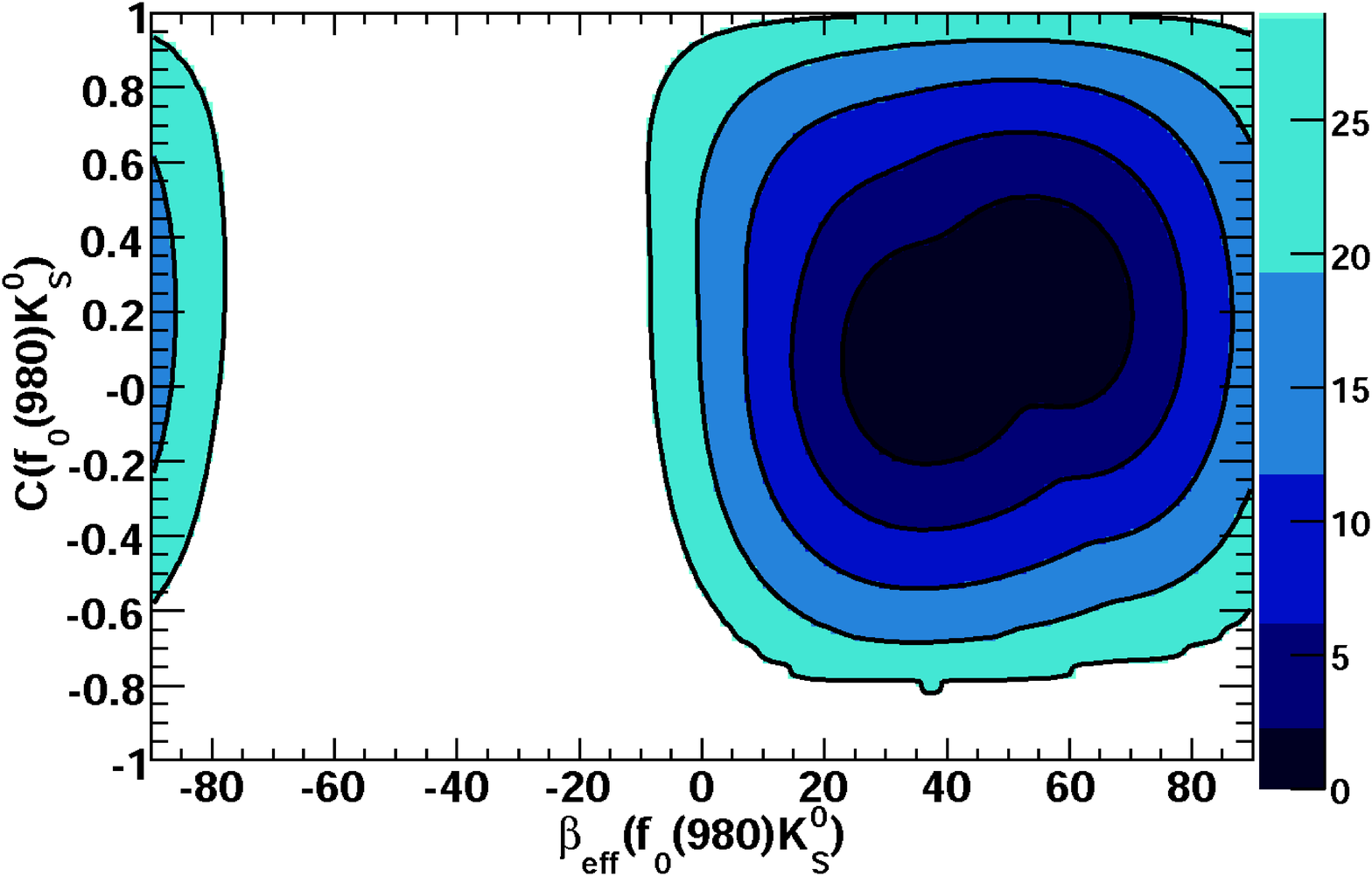} &
\includegraphics[height=5cm]{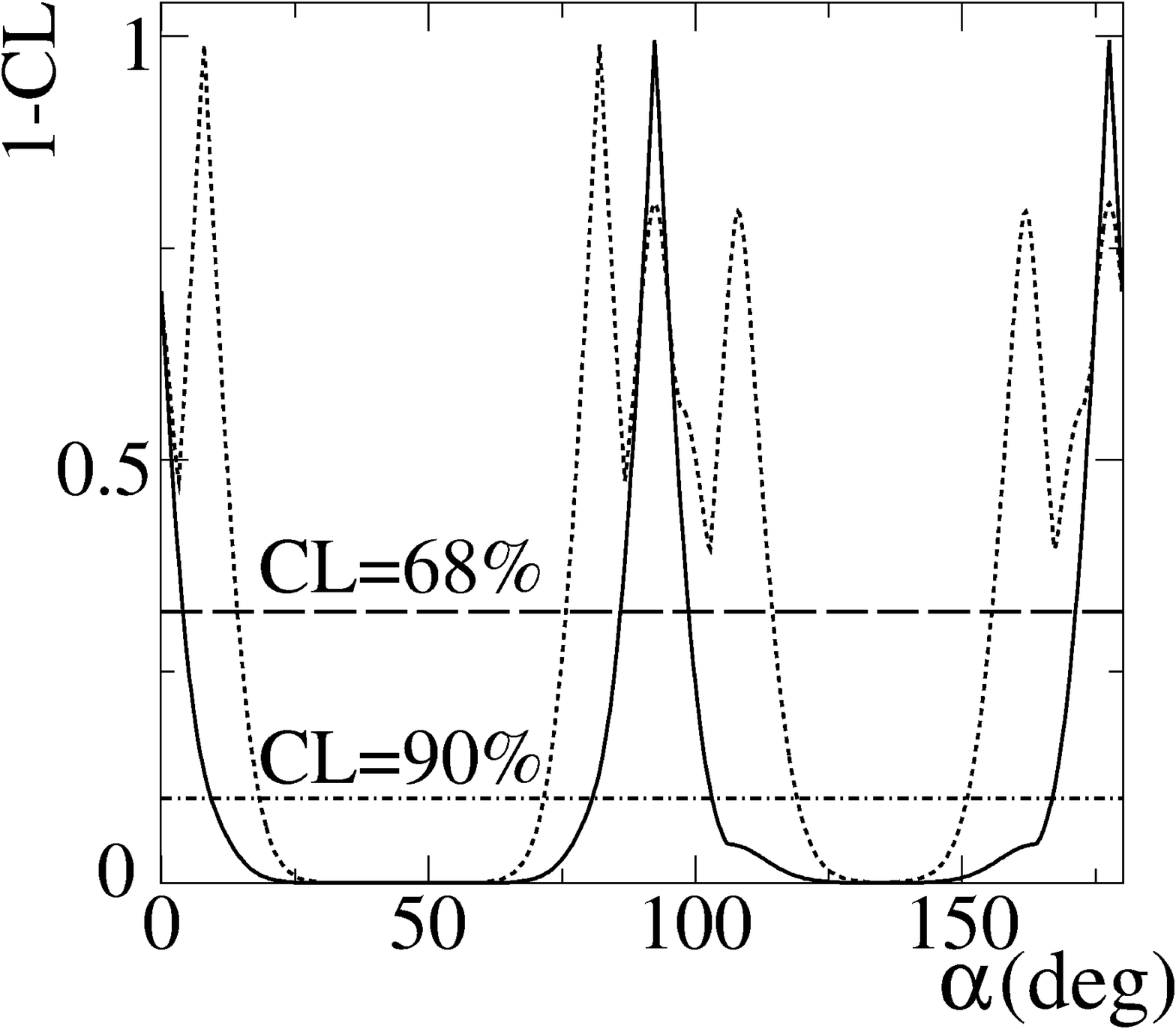} 
\end{tabular}
\caption{\label{fig:angles}Left plot: 1-5$\sigma$ contour plots for the
$f_{0}K^0_S$ mode: on the $x$ axis there is the effective $\beta$ angle, while
on the $y$ axis the term expressing the direct $CP$-violation is represented.
Right plot: constraints on $\alpha$ from the last update of $B^+ \rightarrow
\rho^+ \rho^0$ (solid line); the dotted line represent the constraints
before the inclusion of the last \babar\ result.}
\end{center}
\end{figure}

The final measurement of $\beta$ from a time dependent analysis of the
golden modes $B^0 \rightarrow (c \bar{c}) K^0$ \cite{beta_gold} gives:
\begin{equation}
\sin 2\beta = 0.687 \pm 0.028
\end{equation}
The $\sim3\sigma$ discrepancy between the golden modes and the modes 
dominated by penguin amplitudes which was seen in 2004, has shrunk
considerably, especially in the theoretically cleanest modes, like
$\phi K^0$ and $\eta^{\prime} K^0$. Another determination of $\beta$
has been obtained from a Dalitz Plot analysis of the $B^0 \rightarrow
K^0_S \pi^+ \pi^-$ decay. The two solutions found (see Fig.~\ref{fig:angles})
are in good agreement with the result from the golden modes and we
see evidence of $CP$-violation in the $f_{0}K^0_S$ mode.

The $\alpha$ angle can be extracted, within an 8-fold ambiguity, from an 
isospin analysis of $B \rightarrow \rho^+\rho^-, \rho^0 \rho^0, \rho^+ \rho^0$ 
decays, as proposed in \cite{rho_th}. This analysis was performed after
\babar\ obtained the result on the measurement of the branching fraction
and direct $CP$-violation of $B^+ \rightarrow \rho^+ \rho^0$, based on the
full dataset \cite{rhorho}. The measured branching fraction:
\begin{equation}
{\cal B}(B^+ \rightarrow \rho^+ \rho^0) = (23.7 \pm 1.4 \pm 1.4) \times 10^{-6}
\end{equation}
\textit{flattens} the isospin triangles, thus allowing the removal
of some of the ambiguities with respect to the previous analysis.
The result of the isospin analysis is (discarding the solution close
to zero):
\begin{equation}
\alpha = (92.4^{+6.0}_{-6.5})^{\circ} \, .
\end{equation}

\section{Conclusions}

At the end of the extensive experimental campaign carried on by the
$B$-factories \babar\ and Belle, the $CKM$ mechanism has proven to
be successful in explaining all the Flavour Physics phenomena. The
global fits combining all the measurement relevant for the determination
of the parameters of the $CKM$ matrix show no significant discrepancy
between sets of measurements \cite{fits}. There are some tensions
at the 2$\sigma$ level, for example between the measured value of $\beta$
and the one predicted using the ratio $|V_{ub}|/|V_{cb}|$. Although
not yet significant, the investigation of these discrepancies 
constitute one of the motivations for further pursuing the precision
measurements on Flavour Physics at the hadronic colliders (Tevatron
and LHC) and at the next generation of $e^+e^-$ colliders.

\end{document}